\documentclass[conference,usenatbib]{basi}

\usepackage[T1]{fontenc}
\usepackage[british]{babel}
\usepackage[varg]{txfonts}

\usepackage{rotating}
\usepackage{dcolumn}

\begin{document}
\vspace{-0.4cm}
\title[Optical and X-ray Variability of Blazars]{Optical and X-ray Variability of Blazars}
\vspace{-0.6cm}
\author[A. C. ~Gupta]%
       {A. ~C.~Gupta$^1$\thanks{email: \texttt{acgupta30@gmail.com}}, \\
       $^1$Aryabhatta Research Institute of Observational Sciences (ARIES),
Manora Peak, Nainital -- 263002, India}

\pubyear{2015}
\volume{12}
\pagerange{\pageref{firstpage}--\pageref{lastpage}}
%\status{submitted}
\vspace{-0.6cm}
\date{Received --- ; accepted ---}

\maketitle

\label{firstpage}

\vspace{-0.5cm}

\begin{abstract}

Here we report our recent results of variability studies in optical and X-ray 
bands of three blazars namely 3C 273, PKS 2155 -- 304 and BL Lacertae with XMM-Newton. 
We found large amplitude optical to X-rays variability in 3C 273, and PKS 2155 -- 304 
on year time scale. In 3C 273, we noticed that synchrotron cooling and particle 
acceleration are at work at different epoch of observations. In PKS 2155 -- 304, spectral 
energy distribution from optical to X-ray is fitted with LPPL (log parabolic + power 
law) model. In BL Lacertae, optical flux and degree of polarization were anti-correlated.    

\end{abstract}
\vspace{-0.5cm}
\begin{keywords}
  galaxies: active, individual -- BL Lacertae, PKS 2155--304, 3C 273
\end{keywords}
\vspace{-0.5cm}

\section{Introduction}\label{s:intro}
\vspace{-0.5cm}
BL Lac objects (BLLs) and flat spectrum radio quasars (FSRQs) which belong to radio-loud active 
galactic nuclei (AGNs) are clubbed together and known as blazar. BLLs show featureless optical 
spectra while FSRQs show prominent emission lines in their optical spectra. Blazars emit relativistic 
charged particle jets close to our line of sight ($\leq$ 10$^{\circ}$) (Urry \& Padovani 1995).
They show large amplitude flux variation in the complete electromagnetic (EM) spectrum and
variation is strongly polarized. They show variable flux on diverse timescales e.g. timescales 
ranging from a few tens of minutes to less than a day is known as intra-day variability (IDV)
(e.g. Wagner \& Witzel 1995), or intra-night variability or micro-variability; timescales ranging
from days to several weeks is known as short term variability (STV); and timescales from months to
years is known as long term variability (LTV) (e.g. Gupta et al. 2004).  \\
In the present work, I summarize some of our recent results in optical and X-ray EM bands 
on the blazars: BL Lacertae, 3C 273 and PKS 2155--304. We reported the detailed work on these
blazars in (Gaur et al. 2014, Bhagwan et al. 2014, Kalita et al. 2015).    

%------------------------------------------------------------------------------%
\vspace{-0.5cm}
\section{Observations and Analysis}
\vspace{-0.4cm}
For the blazar BL Lacertae, we have taken optical flux and polarization data using 1.5 m KANATA
telescope in Japan (Gaur et al. 2014). For the blazar PKS 2155-304 and 3C 273, simultaneous 
observations from EPIC/pn in (X-ray) and by OM (Optical Monitor) of XMM-Newton telescope are 
taken. During year 2000 to 2012, there were total 20 such observations for PKS 2155-304 (Bhagwan 
et al. 2014) and 30 observations for 3C 273 (Kalita et al. 2015).   

\vspace{-0.5cm}
\section{Results and Discussion}
\vspace{-0.4cm}
Here I briefly report the results of individual blazars. \\
%\\
{\bf BL Lacertae} ($\alpha_{2000.0}$ = 22h 02m 43.3s, $\delta_{2000.0}$ = +42$^{0}$16$^{'}$40$^{"}$; z = 0.069) \\
Optical V band light curve, degree of polarization, and polarization angle are plotted in 
Fig 1 of (Gaur et al. 2014) from bottom to top, respectively. We found that, in the segment 2 
of the light curve, flux and degree of polarization is anti-correlated (see Fig. 2, Gaur et
al. 2014). Anti-correlation in flux and degree of polarization is unique result and probably
detected for the first time for BL Lacertae. The observed phenomenon can be explained by the
more generalized recent model by Larionov et al. (2013), where small variation in Lorentz factor
by keeping other parameters as fixed may show variety of variation in flux and polarization. \\
%\\
{\bf PKS 2155--304} ($\alpha_{2000.0}$ = 21h 58m 52.1s, $\delta_{2000.0}$ = $-$30$^{0}$13$^{'}$32$^{"}$; z = 0.117) \\
In long term optical to X-ray variability, we found that: (a) variations in all bands on years
timescale with a rms amplitude of $\sim$ 35 -- 45 \%; (b) visual inspection show that optical
and X-ray band light curves are not correlated; (c) variability amplitude decreases from optical
to X-ray bands; (d) optical to X-ray spectral energy distribution is well fitted by log 
parabolic (LP) + power law (PL) i.e. LPPL model (Bhagwan et al. 2014). \\
%\\
{\bf 3C 273} ($\alpha_{2000.0}$ = 12h 29m 06.7s, $\delta_{2000.0}$ = +02$^{0}$03$^{'}$09$^{"}$; z = 0.1575) \\       
In long term optical to X-ray variability, we found that: (a) variations in all bands on years
timescale with a rms amplitude of $\sim$ 68 -- 76 \% in optical/UV and $\sim$ 36 -- 42 \% in X-rays; 
(b) visual inspection show that optical and UV band light curves, soft and hard X-ray bands are well 
correlated but optical/UV are not correlated with X-ray; (c) in hardness ratio vs flux plots, we found
clockwise and anti-clockwise loops at different epochs of observations which show synchrotron cooling
and particle acceleration are at work at different epochs (Kalita et al. 2015).

\end{document}